\documentclass[10pt, conference]{IEEEtran}

\usepackage[colorinlistoftodos]{todonotes}

\usepackage{ifpdf}
\ifpdf
\else

\fi

\usepackage{multirow} 
\usepackage{float} 
\usepackage[flushleft]{threeparttable}  
\usepackage{caption}
\usepackage{paralist}
\usepackage{mathtools}

\usepackage{enumitem}
\usepackage{booktabs}

\usepackage{algorithm}
\usepackage{algpseudocode}
\usepackage{algorithmicx}

\usepackage{breqn} 
\usepackage{esvect} 

\algnewcommand{\LineComment}[1]{\State \(\///\) #1}
\algdef{SE}[DOWHILE]{Do}{doWhile}{\algorithmicdo}[1]{\algorithmicwhile\ #1}

\usepackage{amsmath,amssymb,amsthm} 

\ifCLASSINFOpdf
   \usepackage[pdftex]{graphicx}
   \graphicspath{{../pdf/}{../jpeg/}}
   \DeclareGraphicsExtensions{.pdf,.jpeg,.png}
\else
\fi
\usepackage{array}
\begin{document}
\title{
	An Analytical Model for S-ALOHA Performance Evaluation in M2M Networks
}
\author{
	\IEEEauthorblockN{Qipeng Song, Xavier Lagrange and Loutfi Nuaymi}
	\IEEEauthorblockA{
		Networks Security and Multimedia Department, Telecom Bretagne/IRISA, France\\
		Email: firstname.lastname@telecom-bretagne.eu}
}

\maketitle
\begin{abstract}
The S-ALOHA (i.e. slotted-ALOHA) protocol is recently regaining interest in Lower Power Wide Area Networks (LPWAN) handling M2M traffic. Despite intensive studies since the birth of S-ALOHA, the special features of M2M traffic and requirements highlight the importance of analytical models taking into account performance-affecting factors and giving a thorough performance evaluation. Fulfilling this necessity is the main focus of this paper: we jointly consider the impact of capture effect, diversity of transmit power levels with imperfect power control. We propose a low-complexity but still accurate analytical model capable of evaluating S-ALOHA in terms of packet loss rate, throughput, energy-efficiency and average number of transmissions. The proposed model is able to facilitate dimensioning and design of S-ALOHA based LPWAN. The comparison between simulation and analytical results confirms the accuracy of our proposed model. The design guides about S-ALOHA based LPWAN deduced from our model are: the imperfect power control can be positive with capture effect and and appropriate transmit power diversity strategy. The transmit power diversity strategy should be determined by jointly considering network charges level, power control precision and capture ratio to achieve optimal performance of S-ALOHA.
\end{abstract}
%
\IEEEpeerreviewmaketitle
\section{Introduction}
Machine-Type Communication (MTC) is expected to gain more popularity in the next decade. The research efforts to well accommodate the traffic from MTC can be classified into the following two categories: \begin{inparaenum}[1)]
	\item design from scratch dedicated networks, also called Low Power Wide Area Network (LPWAN);
	\item evolve from existing wireless networks, such LTE-M/NB-IoT networks~\cite{song2016survey}.
\end{inparaenum}
However, the MTC traffic exhibits lots of features different from traditional human-to-human communication: large number of connected terminals, small payload but frequent requests. These features pose big challenges for wireless networks based on resource reservation multiple access protocol and make the LPWAN networks more attractive for MTC~\cite{goursaud2015dedicated}. 

Due to its simplicity and low requirement for terminals, the performance study of  of ALOHA-based protocol is regaining interest in the context of the LPWAN in recent years. The objectives of studies are usually analyzing throughput, packet loss rate, etc. In a radio communication system, factors such as capture effect, diversity of transmit power levels, power control precision, among with others, have a significant impact on ALOHA-like multiple access protocol. 

Lots of researches related to ALOHA-like protocol have been done by taking into account some of the aforementioned factors. Lamaire et al.~\cite{lamaire1998randomization} derive the optimal transmit power distribution under three models: perfect capture model, signal-to-interference threshold with and without Rayleigh fading model. Altman et al.~\cite{altman2005slotted} propose to differentiate transmission priority by using different transmit power and convert it as a game problem. Both~\cite{lamaire1998randomization}\cite{altman2005slotted} ignore the impact of power control error for the transmit power distribution. 
Yang et al.~\cite{yang2012performance} analyze backoff algorithms for LTE Random Access Channel (RACH), which employ ALOHA-like protocol. Nielsen et al.~\cite{nielsen2015tractable} analyze the outage probability for LTE four-steps random access mechanism. Both~\cite{yang2012performance}\cite{nielsen2015tractable} use an analytical model adapted from Bianchi model~\cite{bianchi2000performance}. 
Zozor et al.~\cite{zozor2016time} study collision probability for the pure time-frequency ALOHA access via stochastic geometry approach and calculate the load capacity according to a maximal packet loss rate. 
Goursaud et al.~\cite{goursaud2016random} consider the carrier frequency uncertainty issue and study ALOHA protocol behavior. However \cite{yang2012performance}\cite{nielsen2015tractable}\cite{bianchi2000performance}\cite{zozor2016time}\cite{goursaud2016random} have not taken into account the capture effect and diversity of transmit power.
Bayrakdar et al.~\cite{bayrakdar2016slotted} evaluate the throughput performance of S-ALOHA based cognitive radio network under Rayleigh fading channels with capture effect, but with identical transmit power in each transmission.

As far as we know, few works about S-ALOHA protocol jointly consider the impact of capture effect, power control error, diversity of transmit power levels, and give a multi-criteria performance analysis for M2M environment. In this paper, we propose an analytical model to study the steady-state performance of S-ALOHA including packet loss rate, throughput, energy-efficiency and average number of transmission under two situations: \begin{inparaenum}[1)]
	\item ideal system with perfect power control;
	\item wide-band system with imperfect power control.
\end{inparaenum} 
In the proposed model, the basic idea is to numerically obtain cumulative distribution function (CDF) of total interference from its characteristic function (CF) to calculate the capture probability (thus the transmission failure probability for a single trial). We then use a fixed point analysis to calculate the steady state packet loss rate, throughput, energy-efficiency and average number of transmissions.

The remainder of the paper is organized as follows: Sec.~\ref{sec:system-model} presents the system model. Sec.~\ref{sec:ideal_power_control} is the performance analysis about S-ALOHA in ideal systems, i.e., without power control and fading. Sec.~\ref{sec:imperfect-power-control} takes into account imperfect power control under wide-band system. Sec.~\ref{sec:simulation_result} proves the accuracy of proposed model by simulation and and give some deduced design guidelines. Sec.~\ref{sec:conclusion} concludes this paper.

\section{System model}
\label{sec:system-model}
We consider a single base station. 
Terminals served by this base station employ S-ALOHA protocol to transmit packets. 
The behavior of different terminals is independent.
The time axis is divided into slots of fixed length $T$ equal to the transmission time of a single packet. 
The fresh arrival packet is modeled as a stationary Poisson process with mean intensity $\lambda$. 
Thus, the fresh packet arrival rate in unit slot is $\alpha = \lambda T$. 

In terms of capture effect, Signal-to-Interference-and-Noise ratio (SINR) threshold model is applied. 
In such a model, the other simultaneous packet transmissions are interference sources for a given transmission. 
One packet transmission is failed if and only if its received SINR is less than a threshold $T_{\text{thres}}$ (also named capture ratio), which depends on the modulation coding and receiver characteristics~\cite{dardari2000capacity}. 
In case of transmission failure, the retransmission is scheduled after a random number of slots. 
Each packet is allowed to be retransmitted at most $K$ times. 

Since retransmissions take place at random over long intervals following the collisions that give rise to them and according to Poisson's splitting property~\cite{meyn2012markov}, 
the aggregate packet arrival process can be divided into $K+1$ mutually independent Poisson arrivals processes, each with mean intensity $\alpha P_k$, where $P_k$ is the steady-state probability for a packet to make at least $k$ retransmissions. Note that $P_0$ is always $1$ and $P_{K+1}$ is the steady-state packet loss rate. The steady-state throughput is $\alpha\left( 1 - P_{K+1}\right)$. 

It is difficult for low-cost M2M terminals to support complicated transmit power level diversity strategies. In our model, we employ a simple method to adjust the transmit power: for each retransmission, the transmit power level is multiplied by a factor $v$. According the value of $v$, we define three strategies to be evaluated:\begin{inparaenum}[a)]
	\item identical power level $v=1$; 
	\item power increase with factor $v > 1$;
	\item power decrease with factor $v < 1$.
\end{inparaenum}
One device is assumed to have a valid transmit power interval. For the first and second strategies, the transmit level starts from the minimum valid power level. For the last strategy, the last transmission trial (i.e., for the $K$th retransmission) use the minimum valid power level.
 
Let $p_k$ be the received power at the base station for the $k$th retransmission, due to capture effect, given that the background noise is negligible compared to interfering transmissions, the failure probability $Q_{k}$ of a $k$th retransmission is thus $Q_{k} = Pr\left\lbrace p_k/I < T_{\text{thres}} \right\rbrace$, where $I$ is the cumulative interference caused by all other simultaneous transmissions. Thus, $Q_{k}$ is by nature a function of probability vector $\left\langle P_0, P_1, ..., P_K\right\rangle$. We thus obtain a fixed point equation array between $\left\langle P_0, P_1, ..., P_{K+1}\right\rangle$ and $\left\langle Q_0, Q_1, ..., Q_K\right\rangle$ as follows:   
\begin{align}
\label{eq:recurrisve-array}
	P_0 = 1; \text{ }P_1 = P_0 Q_0; \text{ }...; \text{ } P_{K+1} = P_{K} Q_{K}
\end{align}
Starting with $\left\langle 1, 0, 0, ..., 0\right\rangle $ for $P_0, P_1, ..., P_K, P_{K+1}$, we iteratively obtain the probability vector $\left\langle P_0, P_1, ..., P_K, P_{K+1}\right\rangle$.

Apart from steady-state packet loss rate and throughput, the probability vector $\left\langle P_0, P_1, ..., P_K, P_{K+1}\right\rangle$ allows to analyze average energy efficiency $\overline{\text{EE}}$. The latter is defined as the ratio between number of delivered packets and the total energies consumed (including for dropped packets). For an ergodic stochastic process, statistical average of energy efficiency can be well approximated by its temporal average. Note that what we care is the impact of transmit power diversity and power control error on average energy efficiency. The attenuation caused by propagation distance can be ignored without affecting the performance comparison result. The average energy efficiency thus can be expressed in terms of received powers:
\begin{align}
\overline{\text{EE}} &= \frac{1-P_{K+1}}{\sum_{k=0}^{K} P_{k}\overline{J_k}},
\end{align}
where $\overline{J_k} = p_k T$ is the average energy consumed by a packet on $k$th retransmission. 

The expected number of transmissions $\overline{N_{Tx}}$. This metric allows to give an overview about the average delay of S-ALOHA.
\begin{align*}
	\overline{N_{Tx}} &= \sum_{k=0}^{K} P_{k} 
\end{align*}
In the following, we show how to numerically calculate probability vector of interest under two situations: ideal system without power control error and wide-band system with imperfect power control.
\section{Ideal systems with perfect power control}
\label{sec:ideal_power_control}
In this section we assume the power control is perfect. Fading and shadowing effect is ignored. At the first transmission, all terminals transmit at a power such that the received power at the base station is constant. Let $c_{\mbox{ref}}$ be that power. At each retransmission, the power is multiplied by a factor $v$. Hence, at the $k$th re-transmission (note that $k=0$ for the first transmission), the received power $p_{k}$ is $v^k c_{\text{ref}}$. 

In order to keep received power levels as integer, we assume that $v$ can be expressed $v=l/m$ where $l$ and $m$ are integers, and compute power levels normalized by $c_{\mbox{ref}}/m^K$. Hence, at the $k$th transmission, the normalized received power is $p_k=l^k m^{K-k}$. Its corresponding normalized cumulative interference $Y= \sum_{k=0}^{K} Z_k = \sum_{k=0}^{K} \sum_{j=1}^{N_k} l^k m^{K-k}$, where $Z_k=\sum_{j=1}^{N_k} l^k m^{K-k}$ refers to the normalized cumulative interference from $k$th retransmission Poisson process and $N_k$ denotes the number of packets on $k$th retransmission following Poisson distribution with average arrival rate $\alpha P_k$. 

The cumulative interference component $Z_k$ is a compound random variable~\cite{ross2014introduction}, whose Laplace transform is detailed in Appendix~\ref{annexe:laplace-transform-compound-RV}. Applying ($\ref{eq:laplace-transform-comound-RV}$), we have:
\begin{align*}
\mathcal{L} \left\lbrace Z_k \right\rbrace \left( s \right)
&= \exp\left\lbrace \alpha P_k\left( \exp(-sl^k m^{K-k})-1\right)\right\rbrace, 
\end{align*}
where $\mathcal{L} \left\lbrace f(\cdot) \right\rbrace \left( s \right)$ is the Laplace transform operator with complex variable $s$ for function $f(\cdot)$ .

Since the series of random variables $Z_k, k=0,..., K$ are independent, the Laplace transform of $Y$ is:
\begin{align*}
\mathcal{L} \left\lbrace Y \right\rbrace \left( s \right) 
&= \prod_{m=0}^{K}  \mathcal{L} \left\lbrace Z_k \right\rbrace \left( s \right) \\
&= \exp\left\lbrace \alpha \left( \sum_{k=0}^{K} P_k \exp(-sl^k m^{K-k})-\sum_{k=0}^{K}P_k\right)\right\rbrace
\end{align*}
With a substitution $s= -i\omega$, we obtain the characteristic function $\phi_{Y}\left( \omega \right)$ of $Y$:
\begin{align*}
\phi_{Y}\left( \omega \right) &= \exp\left\lbrace \alpha \left( \sum_{k=0}^{K} P_k \exp(i \omega l^k m^{K-k})-\sum_{m=0}^{K}P_k\right)\right\rbrace
\end{align*}
Note that $Y$ is a discrete random variable. Via a numerical integral method detailed in~\cite{nuttall1969numerical}, the cumulative distribution function $F_{Y}\left( x \right)$ of $Y$ can be derived from its characteristic function $\phi_{Y}\left( \omega \right)$.
\begin{align}
\label{eq:pr_c_m}
F_{Y}\left( x \right)  &= 	\frac{1}{\pi}\int_{0}^{\pi} \frac{\sin\left[ (x+1) \omega/2 \right] }{\sin\left[ \omega/2\right] } \Re\left\lbrace \phi_{Y}\left( \omega \right) e^{-ix\omega/2} \right\rbrace d\omega,
\end{align}
where $\Re\left\lbrace \cdot \right\rbrace$ is operator taking real part of complex number.

Cumulative distribution function $F_{Y}$ can be numerically and rapidly obtained by trapezoidal rule. Due to capture effect, the transmission failure probability $Q_k$ of a packet on $k$th retransmission is thus:
\begin{align}
\label{eq:failure_pb}
	Q_{k} &= Pr\left\lbrace \frac{l^k m^{K-k}}{Y} < T_{\text{thres}}\right\rbrace \nonumber\\
	&= 1 - F_Y\left( \lfloor\frac{l^k m^{K-k}}{T_{\text{thres}}}\rfloor \right),
\end{align}
where operator $\lfloor x \rfloor$ returns back the maximal integer not greater than $x$. 

Substituting ($\ref{eq:pr_c_m}$) and ($\ref{eq:failure_pb}$) into ($\ref{eq:recurrisve-array}$), we get a fixed point equation for probability vector $\left\langle P_1, ..., P_{K+1}\right\rangle$.  

\section{Wide-band Imperfect power control}
\label{sec:imperfect-power-control}
 We consider that data packets are transmitted with a wide-band signal (e.g. by use of a spread spectrum technique). Hence, there is no Rayleigh fading. The received power level of packet transmission is influenced by imperfect power control. The effect of imperfect power control in the literature can be assumed to be a multiplier $\epsilon$ following log-normal distribution~\cite{Lee:1992:MCD:530392}. Still let $c_{ref}$ be the received power at the base station without power control error. Normalized by $c_{ref}$, the received power $p_k$ for the $k$th retransmission, for a given device with index $i$, can be written as follows:
\begin{align*}
p_{ki} = v^k e^{\beta \epsilon_i}, \text{with } \beta = \frac{\ln(10)}{10}
\end{align*}
The power control error factor $\epsilon_i$ is a zero-mean Gaussian random variable with a standard deviation of $\sigma$, namely $\epsilon_i \sim N\left( 0, \sigma^2\right)$. 

The cumulative interference $I$, caused by those terminals simultaneously transmitting packet,  during the $k$th retransmission is thus:
\begin{align*}
I = \sum_{m=0}^{K} v^{m}\sum_{j=1}^{N_m} e^{\beta \epsilon_j},
\end{align*}
where $N_m$ refers to the number of packets on $k$th retransmission and follows Poisson distribution with arrival rate $\alpha P_m$. 

Due to capture effect, a $k$th retransmission trial is failed under the condition that the ratio between received power $p_k$ and cumulative interference $I$ is less than a threshold $T_{\text{thres}}$, namely:
\begin{align*}
\sum_{m=0}^{K} v^{m-k}\sum_{j=1}^{N_m} e^{\beta \left( \epsilon_j - \epsilon_i \right) } &> \frac{1}{T_{\text{thres}}}
\end{align*}

Let focus the normalized cumulative interference $Y_k$ corresponding to a packet on $k$th retransmission:
\begin{align*}
Y_k &=\sum_{m=0}^{K} v^{m-k}\sum_{j=1}^{N_m} e^{\beta \left( \epsilon_j - \epsilon_i \right) } \\
&=\sum_{m=0}^{K}\sum_{j=1}^{N_m} e^{\left(m-k\right) \ln(v)+\beta \left( \epsilon_j - \epsilon_i \right) } 
\end{align*}
With substitution $\theta = \left(m-k\right) \ln(v)+\beta \left( \epsilon_j - \epsilon_i \right)$, $Y_k=\sum_{m=0}^{K} Z_m  = \sum_{m=0}^{K}\sum_{j=1}^{N_m} e^{\theta}$. Since $Z_m$ for $m=0,...,K$ are mutually independent, thus, the Laplace transform of $Y_k$ is:
\begin{align}
	\label{eq:laplace-transform-y-form-1}
	\mathcal{L} \left\lbrace Y_k \right\rbrace \left( s \right) &= \prod_{m=0}^{K} \mathcal{L} \left\lbrace Z_m \right\rbrace \left( s \right) \nonumber\\
	&=\prod_{m=0}^{K} \exp{\alpha P_m \left(  \mathcal{L} \left\lbrace e^{\theta} \right\rbrace \left( s \right)  - 1\right) }
\end{align}
Random variable $\theta$ follows a normal distribution with mean $\left(m-k\right) \ln(v)$ and variance $2\beta^2\sigma^2$. Namely $\theta \sim \mathcal{N}\left( \left(m-k\right) \ln(v), 2\beta^2\sigma^2\right)$.

A closed form expression of the Laplace transform of the lognormal distribution does not
exist. According to reference~\cite{asmussen2016laplace}, the Laplace transform of a log-normal random variable can be approximated as follows:
\begin{align}
\label{eq:laplace-transform-lognormal-form-1}
\mathcal{L} \left\lbrace e^{\theta} \right\rbrace \left( s \right)
&= \frac{\exp(-\frac{W(s \sigma_{\theta}^2 e^{\mu_{\theta}} )^2 + 2W(s \sigma_{\theta}^2 e^{\mu_{\theta}})}{2\sigma_{\theta}^2})}{\sqrt{1 + W(s \sigma_{\theta}^2 e^{\mu_{\theta}})}},
\end{align}
where $W\left( \cdot \right)$ is the Lambert W function~\cite{corless1996lambertw}, which is defined as the solution in principal branch of the
equation $W\left(x\right) e^{W \left( x\right) }= x$.

Combining ($\ref{eq:laplace-transform-y-form-1}$) and ($\ref{eq:laplace-transform-lognormal-form-1}$), with substitution $s = -jw$,  we obtain the characteristic function of cumulative function $\phi_Y\left(w\right)$:
\begin{dmath}
	\mathcal{L}\left[ Y_k\right] = \exp\left\lbrace \alpha\left( \sum_{m=0}^{K} \frac{P_m}{\sqrt{1 + W(j\omega \sigma_{\theta}^2 e^{\mu_{\theta}})}} \cdot \exp( -\frac{W\left( j\omega\sigma_{\theta}^2 e^{\mu_{\theta}}\right)^2  + 2W\left( j\omega \sigma_{\theta}^2e^{\mu_{\theta}}\right)}{2\sigma_{\theta}^2})
	- \sum_{m=0}^{K} P_m \right) \right\rbrace \nonumber,
\end{dmath}
where $e^{\mu_{\theta}} = v^{\left(m-k\right)}, \sigma^2_{\theta} = 2\beta^2\sigma^2$.

As a continuous random variable, the cumulative distribution function $F_{Y_k}\left( x \right)$ of $Y_k$ can be directly derived from its characteristic function $\phi_{Y_k}\left(w\right)$, for example by use of Gil-Pelaez Theorem~\cite{gil1951note}. However, directly using Gil-Pelaez Theorem needs long time. Applying mathematical techniques used in finance domain~\cite{hirsa2012computational}, we seek to calculate the Fourier transform of $e^{-\eta x} F_{Y_k}\left( x \right)$ where term $e^{-\eta x}$ is a damping function with $\eta > 0$. 
\begin{align}
	\label{eq:intermediate_formula_1}
	\int_{-\infty}^{+\infty} e^{iwx} e^{-\eta x} F_{Y_k}\left( x \right) dx = \frac{1}{\eta - iw} \phi_{Y_{k}}\left( \omega +i\eta \right) 
\end{align}
Applying Fourier inversion for ($\ref{eq:intermediate_formula_1}$), we obtain the expression for $F_{Y_k}\left( x \right)$ as follows:
\begin{align}
\label{eq:pr_c_m_case2}
F_{Y_k}\left( x \right)  &= \frac{e^{\eta x}}{2\pi} \int_{-\infty}^{+\infty} e^{-i \omega x} \frac{1}{\eta - i\omega} \phi_{Y_k}\left( \omega +i\eta\right) d\omega  \nonumber\\
&= \frac{e^{\eta x}}{\pi} \Re\left\lbrace  \int_{0}^{+\infty} e^{-i \omega x} \frac{1}{\eta - i\omega} \phi_{Y_k}\left( \omega +i\eta\right) d\omega\right\rbrace, 
\end{align}
The cumulative distribution function $F_{Y_k}\left( x \right)$ now can be derived directly from ($\ref{eq:pr_c_m_case2}$) using a single numerical integration.

The transmission failure probability for the $k$th retransmission $Q_{k}$ is:
\begin{align}
\label{eq:failure_pb_case2}
Q_{k} &= 1- F_{Y}\left( \frac{1}{T_{\text{thres}}} \right) 
\end{align}
Similar with what we do in Section~\ref{sec:ideal_power_control}, combining ($\ref{eq:pr_c_m_case2}$)($\ref{eq:failure_pb_case2}$) and ($\ref{eq:recurrisve-array}$), we can use fixed point method to get the solution for probability vector $\left\langle P_0, P_1, ..., P_K\right\rangle$.

\section{Simulation results and discussion}
\label{sec:simulation_result}
\subsection{Accuracy of proposed models}
To verify the accuracy of proposed analytical model, we develop a python-based simulator. In this simulator, we define $N$ M2M devices. Each device generates a fresh packet with probability of $\alpha/N$ in each slot. The total number of packets generated by all devices during one slot approximately follows a Poisson distribution with intensity $\alpha$ if $\alpha/N$ is enough small. In case of transmission failure, a retransmission is scheduled after a random number of slots following exponential distribution with mean of $36$ slots. For the diversity of transmit power, we consider three strategies: \begin{inparaenum}[1)]
	\item identical transmit power level $v=1$; 
	\item incremented transmit power level with factor $v=2$;
	\item decremented transmit power level with factor $v=0.5$.
\end{inparaenum}
The maximum allowed retransmission number $K$ is set as $4$. In terms of capture effect, we confirm our analytical model under three capture ratios: $3$dB, $0$dB, $-3$dB.  

\begin{figure*}[!tb]
	\centering
	\includegraphics[width=1.0\linewidth]{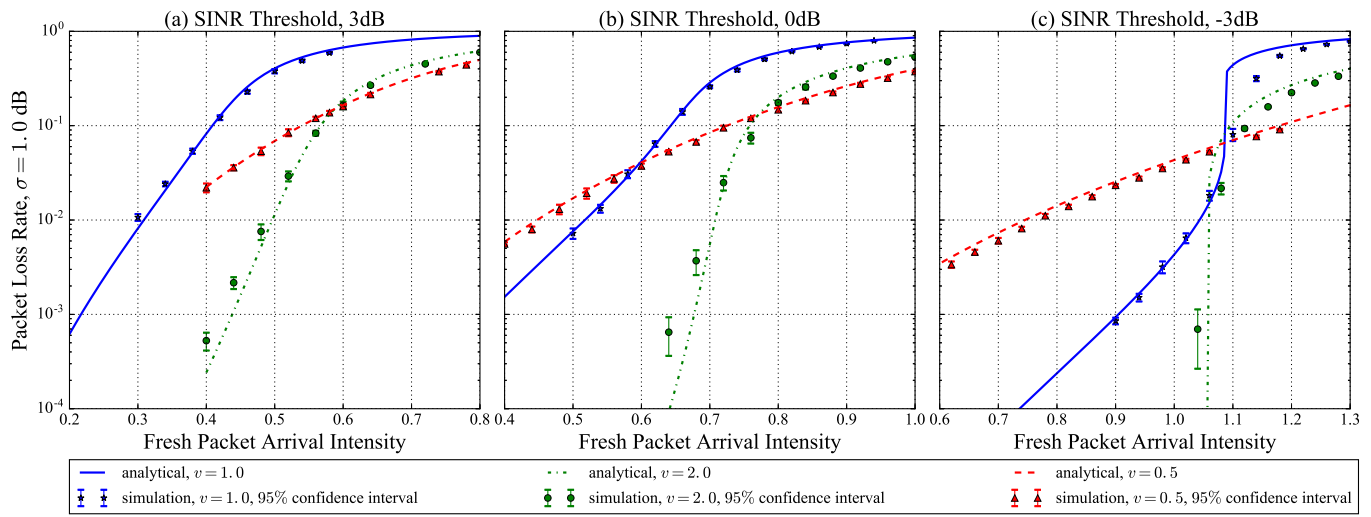}
	\caption{The packet loss rate of S-ALOHA under wide-band system. The power control error is $1$dB.  The power increment factor $v=1, 2, 0.5$ respectively correspond to the case: identical transmit power, incremental transmit power, decremental transmit power. Simulation is repeated $40$ times for each arrival intensity.}
	\label{fig:ci}
\end{figure*}

Due to the limitation of space, just the packet loss rate comparison between analytical and simulation results ($95 \%$ confidence interval) of wide-band system with power control error is shown in Fig.~\ref{fig:ci}. We observe that, the proposed analytical results coincide with that of simulation in most cases. There exists a difference between analytical and simulation result in Fig.~\ref{fig:ci}(c), when fresh packet arrival rate $\alpha$ is between $1.08$ and $1.1$. For regime of interest, from $10^{-3}$ to $10^{-1}$, the proposed models give accurate estimation of packet loss rate.
 
For a given arrival rate $\alpha$, with our proposed analytical models, the probability vector can be obtained within $30$ iterations, within several seconds. This means that the proposed models can be integrated into M2M network dimensioning tools box.
\subsection{Performance evaluation under different settings}
\begin{figure*}[!tb]
	\centering
	\includegraphics[width=1.0\linewidth]{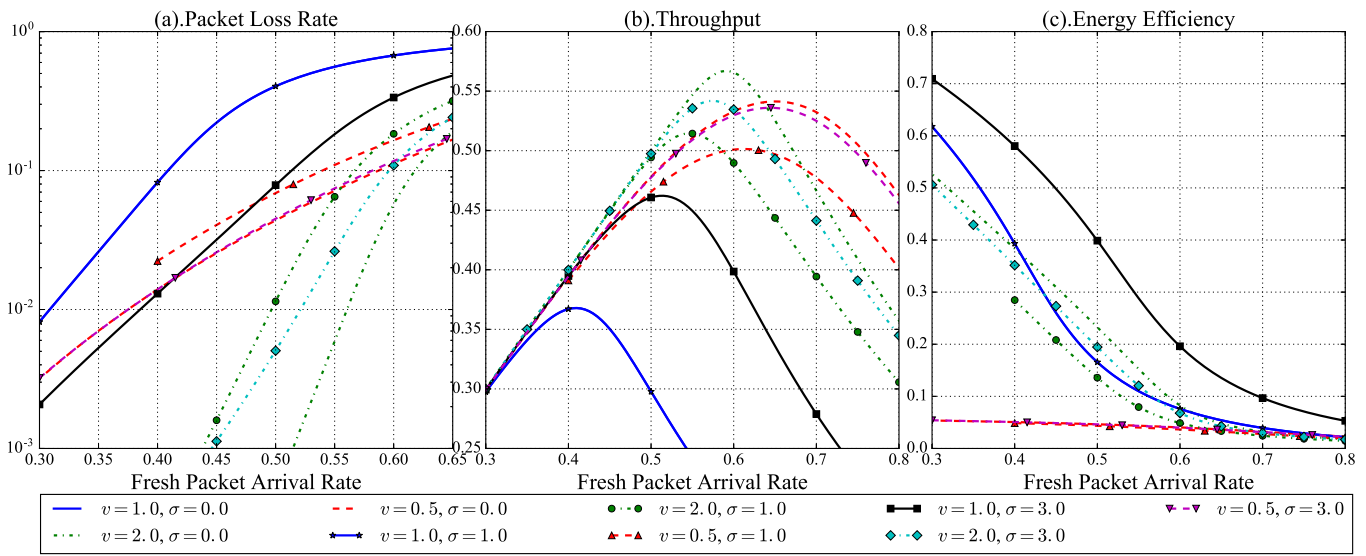}
	\caption{Performance comparison under capture ratio $3$dB. Each sub-figure shows the comparison under three different power control error standard deviation $\sigma = 0, 1, 3$dB. Note that $\sigma=0.0$dB refers to the perfect power control case.}
	\label{fig:shadowing_performance}
\end{figure*}
The S-ALOHA in M2M networks is evaluated with packet loss rate, throughput, energy efficiency. Due to limitation of space, the performance of average number of transmission is not plotted in the figure. We compare the performance under power control error $0$dB (i.e., perfect power control), $1$dB, $3$dB in each figure. The case of perfect power control serves as comparison reference.

In Fig.~\ref{fig:shadowing_performance} with capture ratio $3$dB, we observe that the imperfect power control has positive impact if $v=1$, namely with identical transmit strategy. When power control error standard variance is $1$dB, the performance of S-ALOHA is identical with perfect power control case (the solid line with stars completely superposes with solid line). When the standard deviation of power control error is $3$dB, the S-ALOHA performance with identical transmit strategy gets improved (comparing the solid line with square and solid line with stars). In this case, power control error acts as a way to make transmit power levels more diverse and improve the performance. For other two strategies $v=2$ and $v=0.5$, the imperfect power control degrades the performance of S-ALOHA, because power control error reduces the transmit powers levels diversity introduced by factor $v$. In Fig.~\ref{fig:shadowing_performance}(c), S-ALOHA using identical transmit power outperforms than other strategies in terms of energy-efficiency. The energy-efficiency of decremental power strategy is always at low level, since this strategy requires to start with high power levels. 

\begin{figure*}[!tb]
	\centering
	\includegraphics[width=1.0\linewidth]{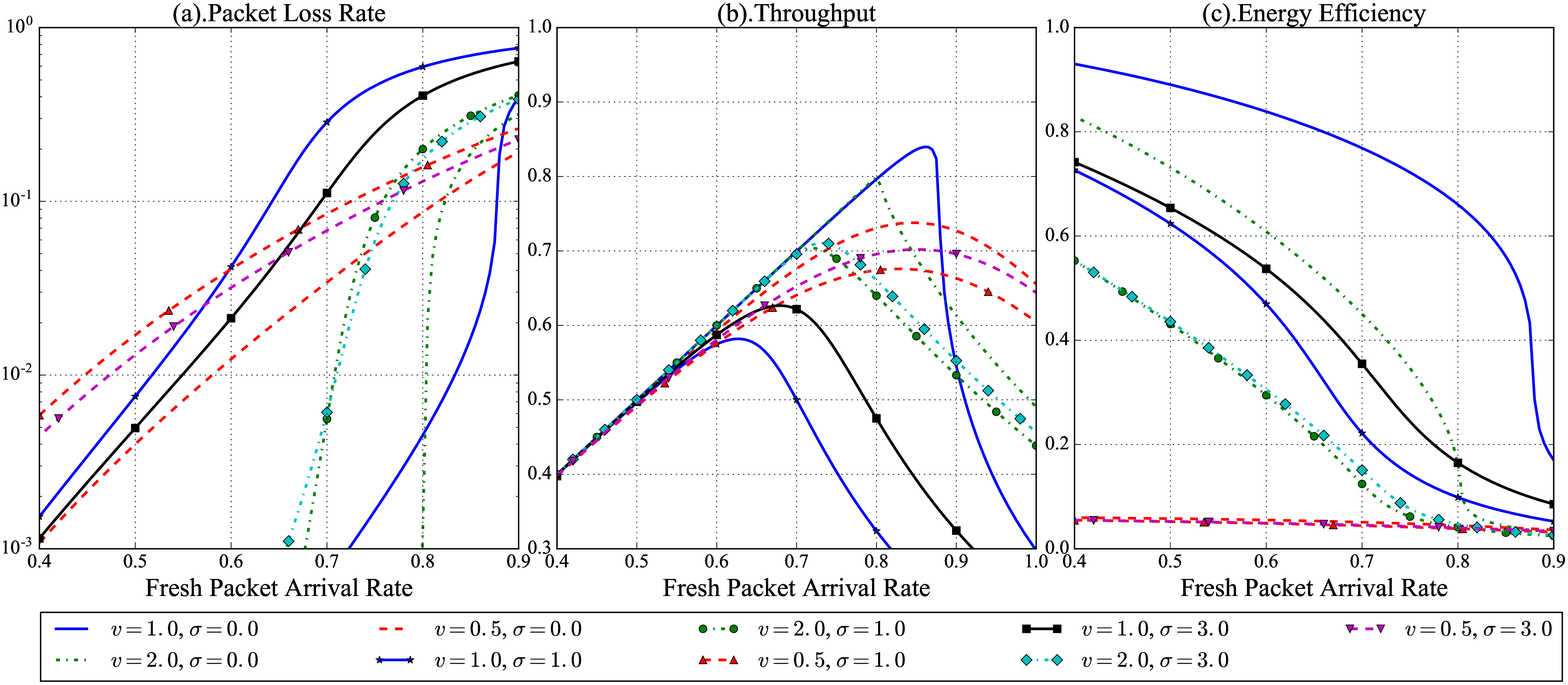}
	\caption{Performance comparison under capture ratio $0$dB. Each sub-figure shows the comparison under three different power control error standard deviation $\sigma = 0, 1, 3$dB. Note that $\sigma=0.0$dB refers to the perfect power control case.}
	\label{fig:shadowing_performance_0}
\end{figure*}

Fig.~\ref{fig:shadowing_performance_0} shows the comparison result under capture ratio $0$dB. In this case, the performance of S-ALOHA gets worse due to power control error no matter which strategy is employed. For all three power strategies, the performance of power control error $3$dB is better than that of $1$dB, since serious power control error makes the transmit power more diverse. With such a capture ratio, the best choice of diversity strategy depends on the privileged performance metric. For example, for a network preferring energy-efficiency than throughout, S-ALOHA with identical is its option. Otherwise, S-ALOHA with incremental power strategy is better. 

\begin{figure*}[!tb]
	\centering
	\includegraphics[width=1.0\linewidth]{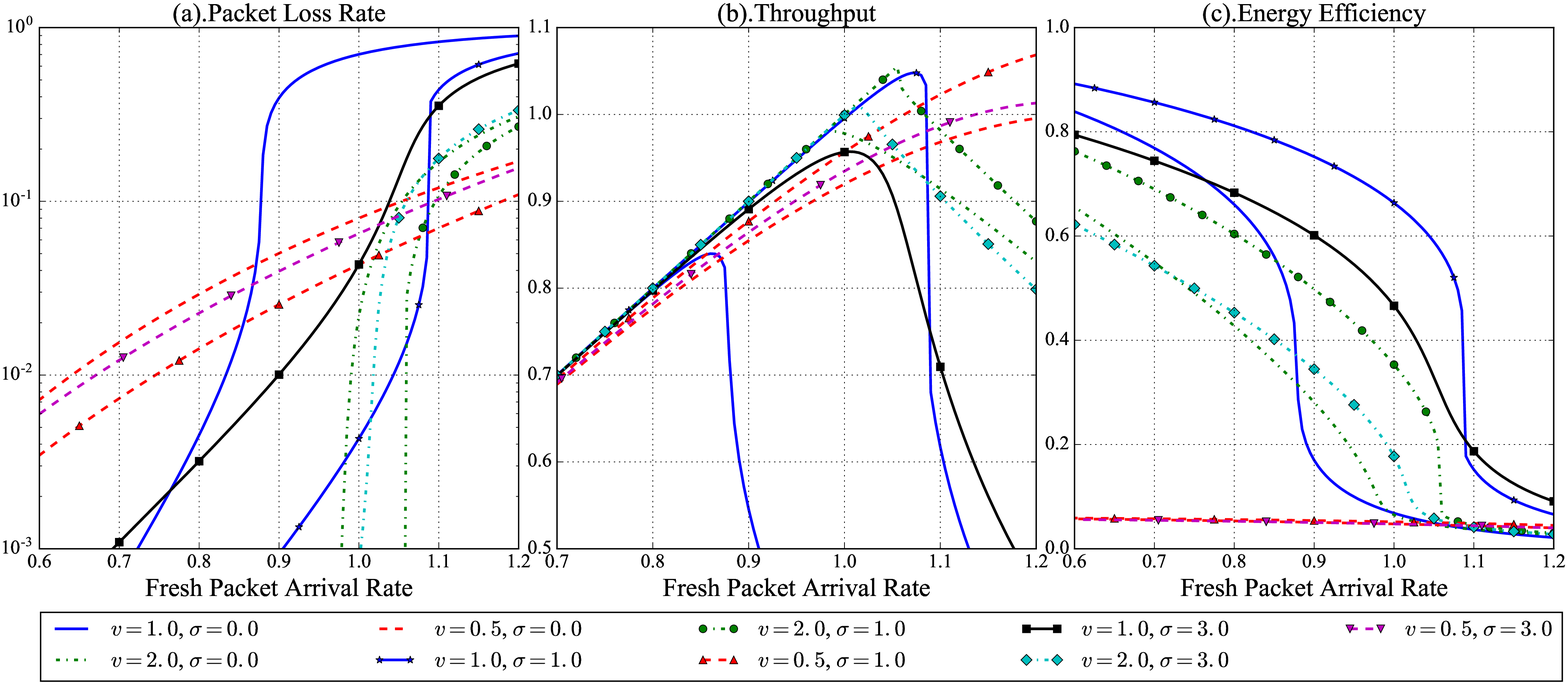}
	\caption{Performance comparison under capture ratio $-3$dB. Each sub-figure shows the comparison under three different power control error standard deviation $\sigma = 0, 1, 3$dB. Note that $\sigma=0.0$dB refers to the perfect power control case.}
	\label{fig:shadowing_performance_-3}
\end{figure*}
Fig.~\ref{fig:shadowing_performance_-3} where capture ratio is $-3$dB refers to a spectrum-spreading system. In this case, the power control error always has a positive impact on S-ALOHA. S-ALOHA achieves a better performance when the power control is more precise. In addition, the choice of transmit power diversity depends on fresh packet arrival rate. For example, when the standard deviation of power control error is $1$dB, if networks based on S-ALOHA are still unsaturated (i.e, with fresh arrival rate less than $1.1$), identical strategy is better than others. With arrival rate greater than $1.1$, the throughput  is sharply reduced. The decremental strategy starts to be a better choice.

\section{Conclusion and outlook}
\label{sec:conclusion}
In this paper we have presented an accurate analytical model capable of estimating steady-state performances, i.e., packet loss rate, throughput, energy efficiency and average number of transmissions, of S-ALOHA based LPWAN networks. The model accounts for various performance-affecting factors, such as capture effect, diversity of transmit power levels, power control error, which have not been jointly considered in previous researches and can not be handled by widely used Bianchi's model.

We employ numerical integration method to calculate cumulative distribution function (CDF) of total inference from its corresponding characteristic function and fixed point analysis to solve the problem. The computational complexity is reduced by combining the recent research effort about log-normal sum (LNS) approximation problem and mathematical skills largely used in finance domain. The accuracy of the proposed model is confirmed by simulation. Due to its low complexity, our model can be used as a dimensioning tool to accurately and rapidly estimate the steady-state system outage capacity and throughput of  S-ALOHA-based LPWAN networks. With our proposed models, we also obtain some design guidelines for S-ALOHA.

In future work, we will add the performance evaluation for narrow-band systems where fading is considered. We will also take into account the impact of  interferences from multiple base stations in the proposed model.
\appendix
\addcontentsline{toc}{section}{Appendices}
\renewcommand{\thesubsection}{\Alph{subsection}}
\label{sec:annexe}
\subsection{Sum of a random number of random variables}
\label{annexe:laplace-transform-compound-RV}
Theorem: a compound random number $S=\sum_{i=1}^{N} X_i$, where $X_i$ are independent identically distributed, $N$ follows Poisson distribution and is independent from $X_i$. Let $S$ be $0$ if $N=0$. The Laplace transform of $X$ is $\mathcal{L} \left\lbrace S \right\rbrace \left( \theta \right) = \exp{\lambda\left(  \mathcal{L} \left\lbrace X \right\rbrace \left( \theta \right) - 1\right) }$.

Proof: The Laplace transform of $S$ is:
\begin{align*}
\mathcal{L} \left\lbrace S \right\rbrace \left( \theta \right)  = \mathbb{E}\left[ e^{-\theta S}  \right] = \sum_{n \geq 0}  \mathbb{E}\left[ e^ {-\theta S }| N = n\right ] \mathbb{P}\left( N = n\right) 
\end{align*}
We have $\mathbb{E}\left[ e^{-\theta S} | N = 0 \right] = 1$, moreover, for $n \geq 1$,
\begin{align*}
\mathbb{E}\left[ e^ {-\theta S }| N = n\right ] 
= \prod_{i=1}^{n} \mathbb{E}\left[ e^ {-\theta X_i}\right] 
= \left( \mathcal{L} \left\lbrace X \right\rbrace \left( \theta \right)   \right) ^ n
\end{align*}
The probability generating function  $G_N\left( z \right)$ of $N$ is:
\begin{align*}
G_N\left( z \right)  = \sum_{n \geq 0} z^n \mathbb{P}\left( N = n \right) 
\end{align*}
With substitution $z=\mathcal{L} \left\lbrace X \right\rbrace \left( \theta \right)$, we have:
\begin{align*}
\mathcal{L} \left\lbrace S \right\rbrace \left( \theta \right)   &= \sum_{n \geq 0} \left( \mathcal{L}_X\left[ \theta \right] \right) ^ n \mathbb{P}\left( N = n \right) 
= G_N\left(  \mathcal{L} \left\lbrace X \right\rbrace \left( \theta \right)  \right) 
\end{align*}
If $N$ follows Poisson distribution with mean $\lambda$, its probability generating function $G_N\left( z \right) =  e^{\lambda\left( z - 1\right) }$. Thus the corresponding Laplace transform is as follows:
\begin{align}
\label{eq:laplace-transform-comound-RV}
\mathcal{L} \left\lbrace S \right\rbrace \left( \theta \right) = \exp{\lambda\left(  \mathcal{L} \left\lbrace X \right\rbrace \left( \theta \right) - 1\right) } 
\end{align}

\bibliographystyle{ieeetr}
\bibliography{new.bib} 

\end{document}